\documentclass[%
 reprint,
superscriptaddress,
 amsmath,amssymb,
 aps,
prl,
]{revtex4-2}

\newcommand{\hmn}[1]{% Hermann-Maguin notation
  \ensuremath{\begingroup\setupHMN #1\endgroup}%
}

\newcommand{\setupHMN}{%
  \doHMN{-}{\HMNoverline}%
  \doHMN{*}{\HMNminverse}%
  \doHMN{i}{\infty}
}

\newcommand{\doHMN}[2]{%
  \begingroup\lccode`~=`#1
  \lowercase{\endgroup\let~}#2%
  \mathcode`#1="8000
}

\newcommand{\HMNminverse}[1]{\frac{#1}{m}}
\newcommand{\HMNoverline}[1]{\mkern1mu\overline{\mkern-1mu#1\mkern-1mu}\mkern1mu}

\usepackage{multirow}
\usepackage[version=3]{mhchem}
\usepackage{svg}
\usepackage{graphicx}% Include figure files
\usepackage{dcolumn}% Align table columns on decimal point
\usepackage{bm}% bold math
\usepackage{xcolor}
\usepackage{float}	
\usepackage{soul}
\usepackage{url}
\usepackage{lipsum}
\usepackage{siunitx}
\usepackage{upgreek}

\DeclareSIUnit{\RPM}{\text{RPM}}
\DeclareSIUnit{\bar}{\text{bar}}
\DeclareSIUnit{\mbar}{\milli\bar}
\DeclareSIUnit{\angstrom}{\text{Å}}
\DeclareSIUnit{\atomicpercent}{\text{at}\%}
\DeclareSIUnit{\ev}{\text{eV}}
\DeclareSIUnit{\mev}{\text{meV}}
\DeclareSIUnit{\kelvin}{\text{K}}
\DeclareSIUnit{\nm}{\text{nm}}
\DeclareSIUnit{\mortsgna}{\text{Å$^{-1}$}}

\begin{document}

\newcommand{\bra}[1]{\left\langle\,#1\,\right|}
\newcommand{\ket}[1]{\left|\,#1\,\right\rangle}

\preprint{APS/123-QED}

\title{Tailoring Emergent Magnetic Moment in La\textsubscript{0.7}Sr\textsubscript{0.3}MnO\textsubscript{3}-Bi\textsubscript{2}Te\textsubscript{3} Heterostructures \textit{via} Interfacial Reconstructions}
%%%%%%%%%%
\author{Damian Brzozowski}
\affiliation
{Department of Materials Science and Engineering, Norwegian University of Science and Technology, Trondheim, Norway}
\author{Yu Liu}
\affiliation
{Department of Materials Science and Engineering, Norwegian University of Science and Technology, Trondheim, Norway}
\author{Øyvind Finnseth}
\affiliation
{Department of Materials Science and Engineering, Norwegian University of Science and Technology, Trondheim, Norway}
\author{Egil Y. Tokle}
\affiliation
{Department of Materials Science and Engineering, Norwegian University of Science and Technology, Trondheim, Norway}
% \author{Karola Neeleman}
% \affiliation
% {Department of Materials Science and Engineering, Norwegian University of Science and Technology, Trondheim, Norway}
% \author{Paul Steadman}
% \affiliation
% {Diamond Light Source, Diamond House, Harwell Science and Innovation Campus, Didcot, Oxfordshire, United Kingdom}
% \author{Christoph Klewe}
% \affiliation
% {Advanced Light Source, Lawrence Berkeley National Laboratory, Berkeley, CA, United States}
% \author{Alpha T. N'Diaye}
% \affiliation
% {Advanced Light Source, Lawrence Berkeley National Laboratory, Berkeley, CA, United States}
% \author{Kenta Amemiya}
% \affiliation
% {Institute of Materials Structure Science, High Energy Accelerator Research Organization (KEK), 1-1 Oho, Tsukuba, Ibaraki, Japan}
% \author{Kaoruho Sakata}
% \affiliation
% {Institute of Materials Structure Science, High Energy Accelerator Research Organization (KEK), 1-1 Oho, Tsukuba, Ibaraki, Japan}
\author{Andrew J. Caruana}
\affiliation
{ISIS Facility, STFC Rutherford Appleton Laboratory, Harwell Science and Innovation Campus, Didcot, Oxfordshire, United Kingdom}
\author{Christy J. Kinane}
\affiliation
{ISIS Facility, STFC Rutherford Appleton Laboratory, Harwell Science and Innovation Campus, Didcot, Oxfordshire, United Kingdom}
\author{Alexander J. Grutter}
\affiliation
{NIST Center for Neutron Research, National Institute of Standards and Technology, Gaithersburg, Maryland, United States}
\author{Dennis G. Meier}
\affiliation
{Department of Materials Science and Engineering, Norwegian University of Science and Technology, Trondheim, Norway}
\affiliation
{Faculty of Physics, University of Duisburg-Essen, Germany}
\author{Ingrid G. Hallsteinsen}
\email{ingrid.hallsteinsen@ntnu.no}
% \phone{+47 735 944 15}
\affiliation
{Department of Materials Science and Engineering, Norwegian University of Science and Technology, Trondheim, Norway}

%%%%%%%%%%%%%%%%%%%%%%%%%%%%%%%%%%%%%%%%%%%%%%%%%%%%%%%%%%%%%
\begin{abstract} 
We report emergent magnetic behavior in heterostructures composed of (111)-oriented La$_{0.7}$Sr$_{0.3}$MnO$_3$ (LSMO) and (00$l$)-oriented Bi$_2$Te$_3$ (BT), controlled by interfacial reconstructions. When BT is deposited directly onto LSMO, an intermediate interfacial layer forms between the two materials. Polarized Neutron Reflectometry modeling reveals that this reconstructed region stabilizes a secondary magnetically ordered phase that is coupled to the underlying ferromagnetic LSMO layer. As a consequence, the heterostructures exhibit unconventional self-crossing magnetic hysteresis loops at room temperature, characterized by a reversal of the net magnetization at low applied fields. In contrast, the introduction of a tellurium seed layer results in a sharper LSMO-BT interface, while preserving the anomalous hysteresis behavior and enhancing the saturation magnetization. Element-specific X-ray absorption spectroscopy suggests that the emergent magnetic phase originates from the chemical reconstruction of manganese species. These results demonstrate that interface engineering in magnetic oxide-topological insulator heterostructures provides a pathway to control emergent magnetic coupling and emergent magnetic states in oxide–topological insulator heterostructures.
\end{abstract}
%%%%%%%%%%%%%%%%%%%%%%%%%%%%%%%%%%%%%%%%%%%%%%%%%%%%%%%%%%%%%
\maketitle
%%%%%%%%%%%%%%%%%%%%%%%%%%%%%%%%%%%%%%%%%%%%%%%%%%%%%%%%%%%%%
Emergent functional properties in material systems have attracted considerable interest for advanced electronic and spintronic applications \cite{KangKT2019,LiuT2020,LiuY2012}. By combining materials with distinct physical characteristics in the form of heterostructure systems, it is possible to enable physical phenomena that cannot be realized in individual constituents alone. In particular, interfaces between materials with dissimilar crystal structures often host novel electronic and magnetic states arising from interfacial reconstruction and symmetry breaking. Such heterostructures have enabled a range of emergent interfacial phenomena, including interfacial superconductivity, controllable Rashba spin-orbit coupling, and axion insulating states \cite{HwangHY2012,NittaJ1997,MogiM2017}.

A major objective is the realization of robust topological transport at elevated temperatures in magnetic topological insulators (TIs), where the interplay between magnetic order and electronic band topology can produce chiral surface conduction channels \cite{ChangCZ2013,KatmisF2016,LiM2017,TokuraY2019}. In the Quantum Anomalous Hall Effect (QAHE), time-reversal symmetry breaking opens an exchange gap in the Dirac surface states, enabling dissipationless edge transport. In practice, however, the temperature at which quantized Hall transport is observed is typically far below both the predicted exchange and the magnetic ordering temperature of the system. This discrepancy arises because the QAHE is highly sensitive to spatial inhomogeneities. Variations in magnetic exchange coupling, fluctuations of the chemical potential, and parasitic bulk or defect conduction channels can all reduce the effective gap controlling quantized transport.

Several approaches have been explored to introduce magnetism into TIs. One widely used strategy involves doping TI chalcogenides with magnetic elements such as V, Cr, or Mn, which has enabled observation of the QAHE at temperatures below a few kelvin \cite{ChangCZ2013,CheckelskyJG2014,MogiM2015,QiS2020}. Alternatively, intrinsically magnetic topological insulators such as \ce{MnBi2Te4} host magnetic order within the topological material itself \cite{OtrokovMM2019,GongY2019,DengY2020,LiQ2024}. However, both approaches suffer from disorder associated with dopant incorporation or structural defects, which can produce spatial fluctuations in the magnetic gap and chemical potential, ultimately suppressing robust quantized transport. An alternative route relies on the magnetic proximity effect (MPE), in which a TI is interfaced with a magnetically ordered material \cite{AlegriaLD2014,TangC2017,ZhuS2018,ZhuS2018_2,PanL2020}. In principle, this approach avoids dopant-induced disorder and enables the use of magnetic materials with high Curie temperatures. In practice, however, realizing strong proximity coupling requires sharp interfaces, low defect densities, and minimal interfacial intermixing. Atomic reconstruction, diffusion, or the formation of intermediate phases can strongly modify the electronic structure at the interface and lead to spatially nonuniform magnetic coupling \cite{RiddifordL2022}. So far, the highest temperature reported for the QAHE in interfacial systems is approximately 0.1 K, far below the magnetic ordering temperatures of the constituent layer \cite{WatanabeR2019}.

In this letter, we examine the FM-TI heterostructures composed of (111)-oriented \ce{La_{0.7}Sr_{0.3}MnO3} (LSMO) and (00$l$)-oriented \ce{Bi2Te3} (BT). LSMO is a ferromagnetic half-metal oxide perovskite with a high Curie temperature, widely studied for its strong correlated electron behavior \cite{TokuraY2000,IzumiM2001} and colossal magnetoresistance \cite{KimuraT1996,BerndtLM2000,LloydJ2017}, while BT is a prototypical model three-dimensional topological insulator \cite{ChenYL2009,ZhangH2009}. Perovskite–chalcogenide FM–TI heterostructures remain largely unexplored. However, several experimental \cite{KorenG2018,Zhu2018,ZhuS2018_2,WangW2019,YangXS2020,LuQ2022,MiaoQ2023} and theoretical \cite{WangW2019,LuQ2022,HuangH2024} reports were presented, including LSMO-based FM-chalcogenide heterostructures \cite{YangXS2020,HuangH2024}. Achieving high-quality interfaces between oxide perovskites and chalcogenide TIs is challenging due to structural and chemical incompatibilities. To address this, we employ two growth approaches: direct deposition of BT onto LSMO, and growth mediated by a tellurium seed layer. The seed layer promotes ordered crystalline nucleation, passivates the surface and acts as a tellurium reservoir during BT growth, thereby reducing interfacial roughness and suppressing the formation of intermediate phases.

The crystalline quality of the samples fabricated is characterized using X-ray diffractometry. Magnetic properties were investigated using bulk magnetometry measurements, as well as element-resolved X-ray absorption spectroscopy and depth-sensitive Polarized Neutron Reflectometry. Depth-resolved magnetic profiles obtained from modeling specular Polarized Neutron Reflectometry data indicate the formation of an intermediate layer between LSMO and BT when the direct-growth approach is employed. In contrast, the data for the seed-layer sample are successively modeled without the need of implementing secondary phases into the stack. In either case, an induced magnetic ordering arises. Magnetic dichroism measurements help determine the contribution of manganese and tellurium species to the interfacial reconstructions and emergent magnetism. Magnetometry measurements reveal unconventional magnetic behavior that is dependent on temperature, FM layer thickness, and the use of seed layers. It is reflected in the formation of self-crossing hysteresis loops and loop shifting along the applied field axis. We further compare these results with a single LSMO thin film and demonstrate that by tuning the thickness of the LSMO layer in LSMO-BT heterostructures, the bulk magnetic response can be modified. By integrating high-temperature FM order with TI layers, this work advances the understanding of magnetic coupling in FM-TI heterostructures and provides design guidelines for robust hybrid heterostructures in future electronic and spintronic applications.
%%% spare parts, scraps, etc. %%%
% The heterostructures were grown to explore emergent magnetic phenomena and elucidate the role of interfacial structure in tailoring magnetic proximity effects. 
% and electron microscopy techniques, complemented by bulk and local structural probes
% and that careful tuning of the BT growth using seed layers strongly affects intermixing and amorphous phase formation at the interface.

\section{Experimental}
\begin{figure*}[hbpt!]
  \centering
  \includegraphics[width=0.8\linewidth]{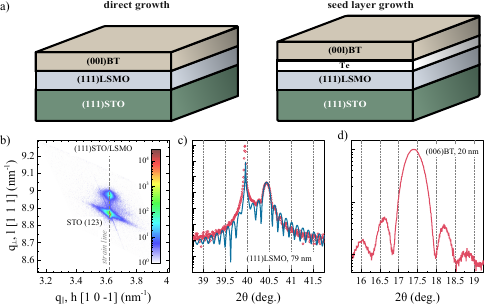}
  \caption{a) Schematic illustration of the LSMO-BT heterostructures discussed. One stack is grown with a direct contact between LSMO and BT layers (left), whereas the other has an additional tellurium seed layer deposited prior to the growth of the BT layer (right). b) RSM of (111)-LSMO around asymmetrical (123)$^+$ peak of STO. The RSM confirms the epitaxial relation between the film and the substrate, showing that LSMO thin film grows under tensile strain on STO. c) HRXRD of (111)-LSMO film deposited onto (111)-STO. The thickness is estimated using the InteractiveXRDFit package. d) HRXRD of (006)-\ce{Bi2Te3} with Te seed layer and estimated 20~nm thick layer.}
  \label{models}
\end{figure*}
LSMO-BT samples were prepared on (111)-oriented \ce{SrTiO3} (STO) substrates (Shinkosha) \textit{via} Pulsed Laser Deposition (PLD) using the direct growth approach and the seed-layer approach with tellurium deposited before the growth of BT. The substrates were cleaned in a sonication bath of acetone, ethanol, and deionized water for 5~minutes each. The cleaning procedure was followed by etching in a buffered hydrofluoric acid solution (1:9 ratio of HF and \ce{NH4F}) for 45~seconds to achieve single termination of the surface. To ensure a smooth step-terrace topography, the substrates were annealed at 1050~°C for 1~hour in \ce{O2} flow.

The substrates were cleaned in an ethanol sonication bath for 3~minutes prior to installation on a PLD holder inside the chamber, and subsequently degassed through the annealing process at 400~°C for 1~hour at a base pressure of $\sim$5$\times$10$^{-8}$~mbar. A KrF excimer laser ($\lambda$~=~248~nm) was used to deposit all layers. For \ce{La_{0.7}Sr_{0.3}MnO3}, the laser was set to the pulsing frequency $\nu$~=~1~Hz and laser fluence $f$~=~2.0~J~cm$^{-2}$. Target to substrate distance $d_{ts}$~=~50~mm. The substrate temperature $T$ was monitored with a pyrometer and set at 650~°C. \ce{O2} gas was introduced into the chamber for the deposition pressure $p$~=~0.3~mbar. Post-deposition annealing was performed at 100~mbar \ce{O2} for 1~hour to minimize oxygen vacancies. Growth conditions of \ce{Bi2Te3} were set as follows: temperature $T$~=~220~°C, pressure $p$~=~1.0~mbar~Ar, target to substrate distance $d_{ts}$~=~45~mm, laser pulsing frequency $\nu$~=~0.2~Hz and fluence $f$~=~0.5~J~cm$^{-2}$. For the seed-layer sample, a thin tellurium film was deposited prior to the deposition of \ce{Bi2Te3}. Tellurium was deposited at room temperature, at a base pressure of $\sim$5$\times$10$^{-8}$~mbar, distance $d_{ts}$~=~50~mm, frequency $\nu$~=~1~Hz, and fluence $f$~=~1.5~J~cm$^{-2}$.

% FEI Apreo scanning electron microscope (SEM) was used to examine morphology and grain coalescence for the first heterostructure. The beam operated at 2.0~kV and 0.2~nA. The image was taken with the back-scattered electron detector. The Transmission Electron Microscopy (TEM) lamella was prepared using a Thermo Fischer FEI Helious G4 UX Focused Ion Beam (FIB) using the lift-out method. The sample was sputter coated with 10~nm of conductive Pt/Pd alloy prior to processing in the FIB. The high-resolution TEM data was acquired on a JEOL ARM200CF at 200~kV, in STO’s (\hmn{10-1}) zone axis. The Scanning TEM–Electron Energy Loss Spectroscopy (EELS) data was acquired on the same instrument at 200~kV, on a GIF Quantum ER. Cross-sectional TEM was used to estimate the thickness of the layers in sample S1, and is presented in Supporting Information.
High-resolution X-ray diffraction (HRXRD) was employed for thickness estimation of each layer. Additionally, reciprocal space mapping was performed around (123)STO peak to assess the lattice straining of LSMO. Both HRXRD and RSM scans were carried out using the D8 Discover diffractometer equipped with a 2-bounce Ge (220) monochromator and Cu K$_{\alpha}$ radiation ($\lambda$~=~1.5406~Å). The HRXRD data of the oxide layers were fitted using the InteractiveXRDFit MATLAB package \cite{interactivexrdfit}.

Depth-resolved Polarized Neutron Reflectometry (PNR) measurements were conducted on thick heterostructures at the PolRef beamline of the ISIS Neutron and Muon Source (Didcot, UK). The measurements were taken at 300~K, under an applied in-plane field of 0.7~T. Open-source \textit{Refl1D} fitting software package was used to build a matching sample model and obtain nuclear and magnetic scattering length density profiles. The considered PNR models were consequently expanded, starting from the simplest stack, and based on the previously collected structural and magnetic data. The final models are discussed in the paper, and the full modeling methodology is presented in detail in the Supporting Information.

Element-resolved X-ray Magnetic Circular Dichroism (XMCD) measurements were performed on manganese L edges and tellurium M edges to confirm the origin of ferromagnetic ordering. The measurements on Mn L$_{3,2}$ edges were performed at the beamline BL-16A of Photon Factory (Tsukuba, Japan), and the measurements on Te M$_{5,4}$ edges were carried out at the beamline I10-1 of Diamond Light Source (Didcot, UK). The data presented were collected using the total electron yield (TEY) mode. An incidence angle was set at 30° relative to the surface. The magnetic field $H$~=~0.7~T was applied parallel to the X-ray beam. The plotted spectroscopy profiles were divided by the $I_0$ beam channel. This is particularly important for the tellurium profiles, as the optics of the beamlines contain chromium, whose L$_{3,2}$ edges are located in close vicinity to tellurium's probed M$_{5,4}$ edges. All profiles were then normalized to the nearest peak's maximum.

A series of magnetometry measurements was performed to characterize the net magnetic behavior of the samples. Because the samples were grown with larger lateral dimensions (10$\times$10~mm$^2$) for PNR measurements, they could not be accommodated in the standard magnetometer sample holder. Therefore, magnetization measurements were performed after the completion of the PNR experiment, after the samples were cut to the appropriate dimensions. The measurements were performed using the Quantum Design Physical Property Measurements System (PPMS) with the Vibrating Sample Magnetometer (VSM) mode. For comparison, additional field-dependent magnetization curves were collected on the single LSMO thin film and an LSMO-BT heterostructure with the thinner LSMO layer using the Superconducting Quantum Interference Device (SQUID) MPMS magnetometer. Prior to each measurement, samples were warmed above their Curie temperature $T_C$, demagnetized using the oscillation mode, and cooled to the target temperature for field-dependent magnetization measurements. Each loop was acquired in the field range of ${\pm}$0.5~T.

\section{Results}
LSMO and BT belong to different crystal systems, rendering high-quality interfaces particularly challenging to achieve. Several reports show the formation of amorphous phases at the interface, which is known to alter the coupling between the materials. One strategy to control the morphology at the interface is to deposit a seed layer prior to \ce{Bi2Te3} deposition, which has been shown to passivate the surface and suppress atomic intermixing. In addition, the tellurium seed layer acts as a tellurium reservoir for newly formed BT layers, minimizing point defects and preserving the 2:3 atomic stoichiometry. With this in mind, two LSMO-BT heterostructures were grown as schematically depicted in Figure~\ref{models}(a)-one with the direct growth, and the other with the seed layer. In order to confirm a high-quality growth of LSMO, RSM is collected around the (123)$^+$ peak of STO. Figure~\ref{models}(b) shows the data collected for the first sample. The RSM is plotted in the scattering vector space, where $q_\parallel$ denotes the in-plane component, and $q_\perp$ the out-of-plane component. The "+" superscript denotes the grazing exit geometry, which in contrast to the grazing incidence geometry probes deeper into the thin film and is more sensitive to structural changes in the bulk of the film. STO and LSMO peaks reside at the same $q_\parallel$ vertical line at 3.62~nm$^{-1}$, confirming that LSMO film is in tensile strain upon growth on (111)-STO. For a relaxed (111)-LSMO film with the pseudocubic lattice constant $a_{pc}$~=~3.87~Å, the (\hmn{123}) peak resides at $q_\parallel$~=~3.65~nm$^{-1}$. This $q_\parallel$ shift then corresponds to the lattice mismatch of -0.9$\%$, in agreement with the expected value. The LSMO peak does not display signs of film relaxation – an otherwise relaxed film would manifest through the peak with a weak intensity tail drifting towards higher $q_\parallel$ values. Figure~\ref{models}(c) depicts high-resolution X-ray diffractograms of the first sample. The corresponding data for the second sample is provided in Supporting Information. The data is fit with the InteractiveXRDFit package. The fitting yields an LSMO epilayer with the thickness of 78.89~nm for the sample with direct growth, and 74.55~nm for the seed-layer sample. Lastly, Figure~\ref{models}(d) shows HRXRD of the (\hmn{006}) peak of BT. The presence of lattice fringes confirms well-defined out-of-plane stacking. From the Bragg's law, the thickness of the BT layer is estimated to be about 20~nm. Thickness calculations are further utilized to construct PNR models. 
% Thickness estimates of the samples were further confirmed by X-ray reflectivity fitting and presented in the Supporting Information.
\begin{table*}
  \caption{Summary of PNR fitting parameters for LSMO-BT heterostructures with the direct growth and seed layer. The samples are split into the corresponding constituents (stack layers), and the thickness and the nuclear/magnetic SLD values are presented. For mSLD, both 300~K and 110~K values are provided in the top and the bottom rows, respectively}
  \label{tbl1}
    \begin{tabular}{c||c|c|c||c|c|c}
    \hline
    \textbf{stack} &  \multicolumn{3}{c ||}{\text{\textbf{direct growth}}} & \multicolumn{3}{c}{\text{\textbf{seed layer}}} \\
    \hline
    & $d$ (nm) & $\rho_N$ (10$^{-4}$~nm$^{-2}$) & $\rho_M$ (10$^{-4}$~nm$^{-2}$) & $d$ (nm) & $\rho_N$ (10$^{-4}$~nm$^{-2}$) & $\rho_M$ (10$^{-4}$~nm$^{-2}$)\\
    \hline
    \hline
    \textit{surface}  & $11.5\pm1.3$ & $1.03\pm0.16$ & 0.00 & $9.2\pm1.0$ & $1.50\pm0.10$ & 0.00\\
    \hline
    \textit{oxidized BT} & $4.7\pm1.3$ & $2.95\pm0.46$ & 0.00 & $5.3\pm1.4$ & $3.15\pm0.54$ & 0.00\\
    \hline
    \multirow{2}{*}{\textit{BT}} & \multirow{2}{*}{$15.4\pm0.9$} & \multirow{2}{*}{$1.73\pm0.09$} & $0.06\pm0.04$ & \multirow{2}{*}{$11.0\pm1.2$} & \multirow{2}{*}{$2.19\pm0.02$} & $0.16\pm0.18$\\
    & & & $0.44\pm0.06$ & & & $0.43\pm0.20$\\
    \hline
    \multirow{2}{*}{\textit{interface}} & \multirow{2}{*}{$10.2\pm2.0$} & \multirow{2}{*}{$3.02\pm0.16$} & $0.85\pm0.09$ & \multirow{2}{*}{--} & \multirow{2}{*}{--} & \multirow{2}{*}{--}\\
    & & & $1.61\pm0.11$ & & & \\
    \hline
    \multirow{2}{*}{\textit{LSMO$_{top}$}} & \multirow{2}{*}{--} & \multirow{2}{*}{--} & \multirow{2}{*}{--} & \multirow{2}{*}{$69.0\pm$} & \multirow{2}{*}{$3.71\pm0.03$} & $1.01\pm0.02$\\
    & & & & & & $1.53\pm0.02$\\
    \hline
    \multirow{2}{*}{\textit{LSMO$_{bottom}$}} & \multirow{2}{*}{$72.9\pm2.1$} & \multirow{2}{*}{$3.81\pm0.02$} & $0.69\pm0.03$ & \multirow{2}{*}{$11.9\pm$} & \multirow{2}{*}{$4.03\pm0.07$} & $1.01\pm0.02$\\
    & & & $1.35\pm0.02$ & & & $1.53\pm0.02$\\
  \end{tabular}
\end{table*}

\subsection{Polarized Neutron Reflectometry}
\begin{figure*}[hbpt!]
    \centering
    \includegraphics[width = \linewidth]{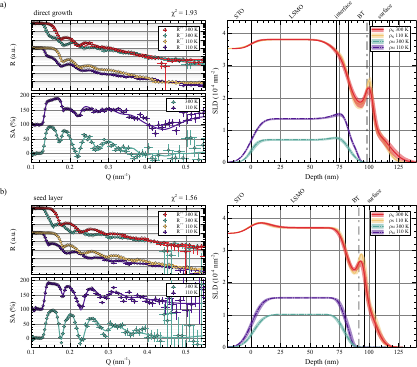}
  \caption{PNR data (reflectivity $R$ and calculated spin asymmetry $SA$) and $\rho_N$/$\rho_M$ profiles of LSMO-BT heterostructures grown using the direct growth (a) and the Te seed layer (b) approach. The data was collected at $T$~=~300~K and 110~K. The 300~K data is shifted along the $y$-axis for clarity. The shaded areas on the SLD profiles mark the 95\% Bayesian confidence intervals. The dashed gray lines indicate the boundary between the pristine and the oxidized BT layers. The achieved figure of merit $\chi^2$ values are written above the plots. For the direct growth sample, the \textit{Refl1D} fit of \(\chi^2=1.93\) predicts formation of an intermediate interfacial phase with induced magnetic moment. For the seed layer sample, the fit of \(\chi^2=1.56\) is achieved without the interfacial layer and with induced magnetic moment within the BT layer.}
  \label{fgrPNR}
\end{figure*}
To provide a nuanced view of the structural-magnetic coupling in the LSMO-BT system, it is critical to address the following questions. Firstly, the influence of Te seed layer on the formation of intermediate amorphous phases must be evaluated. Secondly, a depth-resolved magnetization profile of a sample must be determined to understand whether an emergent magnetic ordering is accommodated at other layers than LSMO. To this end, we performed comparative Polarized Neutron Reflectometry measurements on both samples. The measurements were collected at 300~K and 110~K – slightly above the phase transition temperature of STO $T$~$\approx$~105~K, below which the structure of STO becomes tetragonal. PNR measures the reflected intensity of a spin-polarized neutron beam incident on the sample surface at a grazing angle. It is sensitive to the nuclear ($\rho_N$) and magnetic ($\rho_M$) scattering length density depth profiles of the sample, which are determined by the composition/density and net in-plane magnetization, respectively. By fitting the spin-dependent reflectivities as a function of the momentum transfer vector along the normal direction of the film, the depth profiles $\rho_N$ and $\rho_M$ may be reconstructed.

In addition to spin-dependent neutron reflectivities, it is often instructive to view the spin asymmetry, defined as \(SA=(R^{++}-R^{--})/(R^{++}+R^{--})\), which emphasizes the magnetic contributions to the scattering. Figure \ref{fgrPNR} shows the intensity of the reflected beam $R$, spin asymmetry $SA$, and the nuclear/magnetic scattering length density profiles ($\rho_N$ and $\rho_M$) constructed based on the model fitting using the \textit{Refl1D} software. The detailed modeling protocol is presented in the Supporting Information, where fitting models are built step-by-step from the simplest stack expected. The summary of the sample stacks is shown in Table~\ref{tbl1}.

First, the results of the direct-growth heterostructure are discussed as shown in Figure~\ref{fgrPNR}(a). The optimal fit reaches the figure of merit $\chi^2$~=~1.93. It estimates that the thickness of the LSMO layer is $72.9\pm2.1$~nm, which deviates from the HRXRD fit by approximately -6.0~nm. The roughness at the interface between LSMO and the following layer is $9.0\pm1.0$~nm at 300~K and $8.7\pm1.1$~nm at 110~K. An important feature emerges above the LSMO layer, where an interfacial layer is introduced before BT. The estimated thickness of this layer is about $10.2\pm2.0$~nm. The BT layer has a thickness of $15.4\pm0.9$~nm, and is followed by a subsequent oxidized BT, with a thickness of $4.7\pm1.3$~nm. The sum of the thicknesses of the pristine and oxidized BT layers results in a close match to the HRXRD data. The topmost layer named \textit{surface} refers to various contaminants (commonly hydrocarbon compounds) accumulated on the surface of the samples, which affect the reflectivity decay. Critically, the model predicts that the magnetic moment extends beyond LSMO and reaches both the interfacial layer and the BT layer. The $\rho_N$ values achieved at both temperatures are summarized in Table~\ref{tbl1}. The magnetization at the interface is enhanced with respect to the magnetization of LSMO, which may indicate that manganese species diffuse and accumulate at this layer. It is important to emphasize that alternative models with the fixed zero $\rho_N$ at the interface and BT resulted in less optimal fits. The model constructed for the seed-layer sample shown in Figure~\ref{fgrPNR}(b) yields the figure of merit $\chi^2$~=~1.56 without the interfacial layer and is the most optimal fit achieved for this sample. The other models tested include the introduction of magnetic and non-magnetic interfacial layers, as well as the separate tellurium layers. The model shows that the sample hosts the magnetization higher than the direct-growth heterostructure. Moreover, the magnetization persists within the BT layer, suggesting a partial incorporation of magnetic species within the BT layer or emergent magnetization through the proximity effect. These two mechanisms cannot be distinguished unambiguously from PNR alone. In addition to the absence of an intermediate layer, the roughness of LSMO with the neighboring layer is decreased and equals $5.4\pm0.4$~nm at 300~K and $4.4\pm0.3$~nm at 110~K. The interpretation of the PNR data is that in both the case of the direct growth and the seed layer growth there is an spatially extended magnetization outside of LSMO, and that the seed layer approach provides a more uniform interface between LSMO and BT.

\subsection{X-ray Magnetic Circular Dichroism}
\begin{figure}
  \includegraphics[width=1\linewidth]{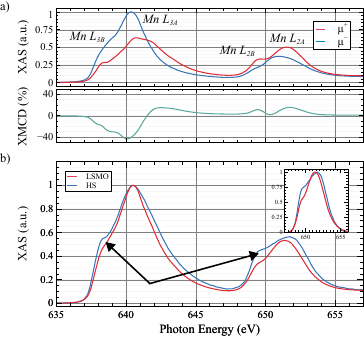}
  \caption{X-ray absorption and dichroism spectra of the Mn L$_{3,2}$ edge. a) XAS and XMCD of the heterostructure with the direct growth. The dichroism intensity is expressed as a fraction of the XAS signal, normalized to the edge's maximum. b) Comparative XAS of the direct-growth heterostructure and a single LSMO thin film. The plot inset shows the Mn L$_{2}$ edge.}
  \label{xas01}
\end{figure}

\begin{figure}
  \includegraphics[width=1\linewidth]{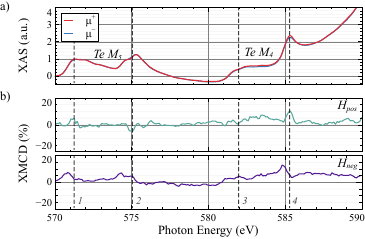}
  \caption{a) X-ray absorption spectra of the Te M$_{5,4}$. b) The corresponding dichroism spectra (green line) and the dichroism spectra collected upon the magnetic field reversal (purple line).}
  \label{xas02}
\end{figure}

The emergence of an intermediate layer between LSMO and BT raises the question of what is the element-specific origin of the stabilized phase and the magnetic ordering. The most likely contributors would be manganese acting as the magnetic reservoir and tellurium as it is highly prone to the desorption and diffusion effects, resulting in Te-deficient BT. Interfacial reconstructions are expected to alter the valency of contributing elements and therefore it is likely that the change can be observed with X-ray absorption spectroscopy. For this reason, XAS and XMCD measurements were carried out on the direct-growth heterostructure. The Mn L$_{3,2}$ edges of LSMO are well-explained in the literature, and generally yield a high signal-to-noise ratio. To compare the Mn edges of the LSMO-BT heterostructure with unaltered Mn edges of a single LSMO layer, comparative analysis is performed and presented in Figure~\ref{xas01}. Figure~\ref{xas01}(a) shows the absorption spectra of Mn L$_{3,2}$ edges of the direct-growth heterostructure. The L$_3$ and L$_2$ edges correspond to the $2p_{3/2}\rightarrow3d$ and $2p_{1/2}\rightarrow3d$ excitations, respectively. Subscripts $A$ and $B$ refer to excitations to the $3d~e_g$ and $3d~t_{2g}$ bands, respectively. The dichroism curve confirms the FM ordering of manganese within the sample, and reaches 42\% at photon energy of 640.4~eV. Figure~\ref{xas01}(b) compares the XAS of the direct-growth heterostructure and a single LSMO film. The LSMO curve (red line) is in agreement with other reports on the Mn$^{3.3+}$ valence state \cite{KavichJJ2007,HallsteinsenI2016}. The heterostructure curve (blue line) deviates from that of a single LSMO layer. An increased intensity is observed at $E$~=~638~eV and 649~eV, marked with arrows, matching the energy of excitations to the Mn $3d$ $t_{2g}$ orbitals. Alternatively, the change may be from a contribution from a lower oxidation state, e.g., \ce{Mn^2+}. The intensity on the high-energy-side is increased, as well. The comparison suggests that manganese may diffuse from the LSMO layer to the intermediate layer, where it stabilizes a phase over an altered oxidation state. This scenario is reinforced under the consideration that the PNR modeling magnetizes the interfacial region.

Unlike manganese, whose L$_{3,2}$ edges yield a robust signal, the tellurium M$_{5,4}$ edges have a weak intensity relative to the background level. In addition, many X-ray magnetic dichroism beamlines employ chromium optics along the pathway, whose L$_{3,2}$ edges lie closely to the tellurium M edges. These challenges make the dichroism analysis of tellurium particularly difficult. Nevertheless, several reports successfully present magnetic dichroism on tellurium edges due to magnetic proximity effects. In order to examine the contribution of tellurium to the interfacial magnetic state, XAS and XMCD are taken on the M$_{5,4}$ edges of the direct-growth heterostructure as shown in Figure~\ref{xas02}. Figure~\ref{xas02}(a) displays two regions of interest: the M$_5$ edge of the $3d_{5/2}\rightarrow5p$ excitation, and the M$_4$ edge of the $3d_{3/2}\rightarrow5p$ excitation. Each region consists of two distinct features numbered at the bottom of the figure and marked with dashed lines, located at 571.2~eV (1) and 575.1~eV (2) for M$_5$, as well as 582.1~eV (3) and 585.3~eV (4) for M$_4$. The peak located at 571.2~eV and the broad feature 582.1~eV (lines 1 and 3) match the tellurium oxidation state expected in \ce{Bi2Te3}. The features at 575.1~eV and 585.3~eV (lines 2 and 4) resemble those originating from the Te$^{4+}$ states present in oxidized tellurium layers such as \ce{TeO2}. Considering that the spectra are acquired in the TEY mode, which is a surface sensitive technique, it is likely that both the surface oxidized states and pristine \ce{Bi2Te3} states are probed at the same time. In addition, it is possible that a weak contribution from chromium, present in the beamline's optics, is seen at 576~eV despite dividing the signal by the total beam intensity $I_0$.

Figure~\ref{xas02}(b) shows the corresponding dichroism curve (green line). At energies corresponding to M$_5$ peaks, there is the dichroism of 5$\%$. The dichroism at the peak located at 585.3~eV (M$_4$ edge, line 4) reaches up to 10$\%$. It must be taken into account that the observed XMCD signal may arise due to artifacts and overall weak signal of the tellurium M edges. Importantly, the measurement taken with the reversed magnetic field (purple line) does not lead to signal inversion at 571.2~eV (1) and 585.3~eV (4), putting in doubt the magnetic contribution of tellurium at these edges. The dichroism at 575.1~eV exhibits sign reversal of about the same magnitude, which may indicate magnetic ordering in the oxidized tellurium phase. However, considering the weak signal of the edges probed, this observation is not conclusive. Conversely, the lack of dichroic effect at tellurium edges does not imply an absence of the magnetic proximity effect. In studies on \ce{Bi2Te3} films interfaced with magnetic layers, the anomalous Hall effect was observed through transport measurements despite no XMCD on the Te M edges. This can be a result of weak bonding between TI and FM, which leads to only a weak transfer of the magnetic exchange. As calculated, the band gap opening of 10~meV led only to a difference of about 1.6$\%$, which lies below the detection limit in our case.

\subsection{Bulk Magnetometry}
\begin{figure*}[hbpt!]
    \centering
    \includegraphics[width = 0.7\linewidth]{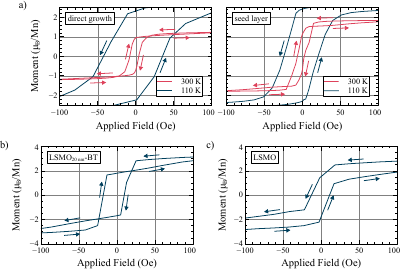}
  \caption{Zoomed-in hysteresis loops around $H$~=~$\pm$100~Oe. a) LSMO-BT heterostructures with the direct growth (left) and seed layer growth (right). Scans collected at two temperatures are plotted to compare the magnetic response. b) Comparative heterostructure with the thinner LSMO layer. c) Single LSMO layer.}
  \label{fgrsquid}
\end{figure*}

Lastly, to evaluate whether interfacial magnetic reconstructions produce a measurable impact on the macroscopic net magnetization, magnetometry measurements were performed. In general, bulk magnetometry techniques such as VSM and SQUID exhibit limited sensitivity to weak induced magnetic moments, such as those associated with magnetic topological insulating phases. Nevertheless, several previous studies have reported modifications in hysteresis behavior and enhanced magnetic responses, as evidenced by $M(H)$ and $M(T)$ measurements.

Figure~\ref{fgrsquid}(a) presents $M(H)$ hysteresis loops for the direct growth and seed-layer heterostructures, measured at $T$~=~300~K and 110~K. The data are magnified near zero field to emphasize low-field features, while full sweep loops are provided in the Supporting Information. At room temperature (red lines), both samples exhibit self-crossing hysteresis loops. Arrows indicate the direction of the magnetic field sweep for individual curves. At high applied fields, the response is characteristic of a conventional ferromagnet, with the magnetization approaching saturation. The normalized saturation magnetization at $T$~=~300~K is 1.34~$\upmu_B$/Mn for the direct growth sample. For the seed-layer sample, the saturation magnetization is enhanced and equals 1.85~$\upmu_B$/Mn.

As the applied field decreases to approximately 10-20~Oe, a pronounced reduction in magnetization occurs, and the magnetic moment reverses direction before the applied field reaches zero. Upon increasing the field in the opposite direction, the magnetization recovers its saturation value. The magnetization reversal process is reproducible in the reverse sweep. Notably, the hysteresis loops are shifted along the applied field axis by approximately $-1.7$~Oe and $-1.9$~Oe for the direct growth and seed-layer samples, respectively. No detectable shift is observed along the magnetization axis.

In contrast, the hysteresis loops measured at $T$~=~110~K (blue lines) do not display self-crossing behavior, but instead exhibit conventional ferromagnetic hysteresis. However, the horizontal shift of the loops remains, indicating that the underlying interfacial coupling persists at lower temperature.

The self-crossing hysteresis loops are not typically observed in standalone (111)LSMO thin films that are the only native magnetic phase deposited on the samples discussed. Such characteristics are commonly attributed to coupled magnetic systems, where two or more interacting magnetic phases give rise to modified net magnetic behavior. The observation of such behavior within the LSMO-BT therefore suggests the possible emergence of an additional magnetic phase beyond the intrinsic ferromagnetism of LSMO.

To elucidate the origin of this anomalous magnetic response, two considerations are addressed. First, the disappearance of self-crossing loops at low temperature may result from the enhanced contribution of the FM LSMO layer to the overall magnetization, effectively masking the coupling effects. This hypothesis can be examined by measuring a similar heterostructure in which the LSMO thickness is substantially reduced, while all other stack components remain unchanged. Second, comparative measurements on a standalone (111)LSMO thin film are required to exclude potential experimental artifacts.

To this end, a comparison is performed with the direct growth LSMO-BT heterostructure having a 20~nm thick LSMO, and with a single LSMO thin film without the BT overlayer. The corresponding hysteresis loops are shown in Figure~\ref{fgrsquid}(b,c). Measurements performed on the heterostructure at 110~K (Figure~\ref{fgrsquid}(b)) confirm the presence of self-crossing loops. Hysteresis curves collected at other temperatures are appended in the Supporting Information. Notably, the behavior persists at the investigated temperature, as well as lower temperatures down to 2~K, supporting the thickness-dependence hypothesis. A loop shift of $-4$~Oe is observed along the applied field axis, without a measurable shift along the magnetization axis. The saturation magnetization $T$~=~110~K is 3.52~$\upmu_B$/Mn. In contrast, the standalone LSMO thin film (Figure~\ref{fgrsquid}(c)) exhibits conventional ferromagnetic hysteresis loops, with no evidence of self-crossing behavior or horizontal loop shift.

\subsection{Discussion}
Magnetic characterization using PNR, XMCD, and magnetometry consistently indicates that interfacial reconstruction plays a central role in determining the magnetic properties of LSMO–BT heterostructures. In the direct-growth case, the formation of an intermediate interfacial layer coincides with the emergence of a spatially extended magnetic moment beyond the LSMO layer. XAS measurements further suggest that this region is associated with modified manganese oxidation states, consistent with the formation of Mn-based secondary phases driven by interfacial diffusion and chemical reconstruction. In contrast, the introduction of a Te seed layer suppresses the formation of a distinct interfacial phase and reduces structural roughness, while still preserving an extended magnetic profile. The persistence of magnetization beyond the LSMO layer in both growth modes indicates that interfacial reconstruction, rather than purely structural sharpness, governs the magnetic coupling in these systems.
\begin{figure}
  \includegraphics[width=0.75\linewidth]{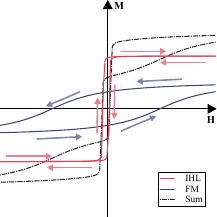}
  \caption{Deconvoluted model of the self-crossing hysteresis loops observed in LSMO–BT heterostructures. The anomalous self-crossing behavior and magnetization reversal are reproduced by superposing two magnetic contributions: a conventional ferromagnetic (FM) component with lower saturation magnetization and higher coercivity, and an inverted hysteresis loop (IHL) component with higher saturation magnetization and lower coercivity.}
  \label{hystsum}
\end{figure}
% Importantly, the introduction of tellurium seed layer prior to the growth of BT profoundly affects the interfacial roughness and, by extension, atomic interdiffusion. While both samples exhibit a similar magnetization characteristics (the self-crossing loops and the shift along the applied field axis at room temperature), the seed-layer heterostructure yields saturation magnetization enhancement relative to the direct-growth heterostructure.

Importantly, the XMCD results do not provide conclusive evidence for a significant magnetic moment on tellurium, suggesting that magnetic proximity effects within the BT layer are not the dominant mechanism responsible for the observed behavior. Instead, the combined data support a scenario in which an emergent manganese-containing interfacial phase contributes an additional magnetic component.

The macroscopic manifestation of this coupling is observed in the form of self-crossing hysteresis loops. These loops can be qualitatively understood as arising from the superposition of two magnetic contributions as schematically deconvoluted in Figure~\ref{hystsum}: a conventional FM LSMO response and a secondary magnetic phase of lower coercivity and antiparallel alignment, resulting in an inverted hysteresis loop characteristic. Such behavior has been reported in coupled dual-ferromagnetic systems and is consistent with an exchange-spring-like mechanism, in which two magnetic phases with different switching fields are exchange-coupled.

While alternative explanations, including experimental artifacts, must be carefully considered, several observations argue against this possibility: (i) the self-crossing behavior is observed exclusively in heterostructures and not in standalone LSMO films; (ii) the measurements are reproducible across multiple magnetometry platforms, including SQUID and VSM systems; (iii) the applied field range is limited, minimizing remanent field effects; and (iv) the magnetic response systematically depends on heterostructure design \cite{OSheaMJ1994,ValvidaresSM2001,GhisingP2020,BhardwajRG2025,WeiG2026}. Taken together, these results support the intrinsic origin of the observed magnetic behavior.

The precise microscopic structure of the interfacial magnetic phase remains an open question. Possible candidates include manganese oxides and manganese–tellurium compounds such as \ce{MnO2}, \ce{Mn3O4}, \ce{MnTe}, or \ce{MnTe2}, all of which can exhibit magnetic ordering \cite{TsengLT2015,VazquezA2005,GonzalezRD2024,ChenW2020}. Further studies combining atomic-resolution structural characterization with element-specific magnetic probes will be required to fully resolve the nature of this phase. Nevertheless, the present results clearly establish that interfacial chemical reconstruction provides a powerful handle to engineer magnetic coupling in oxide–topological insulator heterostructures.

\section{Conclusions}
In summary, we have demonstrated that magnetic properties of LSMO–BT heterostructures can be controlled through interfacial reconstruction. Direct growth of BT on LSMO leads to the formation of an intermediate manganese-containing interfacial phase, while the introduction of a Te seed layer suppresses this phase and improves interface quality. In both cases, Polarized Neutron Reflectometry reveals a spatially extended magnetic moment beyond the LSMO layer.

The presence of an additional magnetic phase gives rise to unconventional self-crossing hysteresis loops, which we attribute to exchange coupling between the LSMO layer and an emergent interfacial magnetic phase. X-ray absorption spectroscopy indicates that this phase is associated with modified manganese chemistry, while magnetic proximity effects in the BT layer appear to play a secondary role.

These results highlight interfacial chemical reconstruction as a key mechanism for tailoring magnetic coupling in oxide–topological insulator heterostructures. These results position perovskite–chalcogenide heterostructures as a compelling platform for the study and control of emergent interfacial magnetism.
% To summarize, we have investigated magnetic properties of LSMO-BT heterostructures through control \textit{via} interfacial reconstructions. By deposition of the intermediate seed layer, the degree of atomic intermixing can be tuned, altering the formation of secondary interfacial phases. PNR modeling shows that an emergent magnetic moment is hosted outside of the FM (111)LSMO. The coupling between magnetic phases leads to magnetic reversal at low applied fields, as evidenced by the magnetometry data. Characteristics of these magnetic hysteresis loops match exchange spring observed in coupled systems of hard and soft ferromagnetic phases. The observations presented in this report provide insight into magnetic perovskite-topological insulating chalcogenide heterostructures, and emphasize on the importance of structural reconstructions at the interface for precise control of magnetic properties.\newline

\begin{acknowledgements}
\section{Acknowledgements}
This project was supported by the Research Council of Norway under Project Number 325063 and the COST Action CA-20116 Network for Innovative and Advanced Epitaxy (OPERA) Short-Term Scientific Mission (STSM). D.B. acknowledges DTU Energy PLD Group (Lyngby, Denmark): Dae-Sung Park, Alessandro Palliotto, Eric Brand, Felix Trier, and Nini Pryds for assistance in film growth and the STSM collaboration; University of Twente (Enschede, the Netherlands), Romar Avila, Emma van der Minne, and Gertjan Koster for assistance in film growth; Gavin Stenning of the Materials Characterisation Laboratory R53 of ISIS (Didcot, UK), as well as Takashi Honda of the High Energy Accelerator Research Organization (Tsukuba, Japan) for their assistance in SQUID/VSM magnetic characterization.
\end{acknowledgements}

% \begin{suppinfo}

% \end{suppinfo}

\bibliography{Bibliography}

@article{Zhu2018,
   author = {Shanna Zhu and Gang Shi and Peng Zhao and Dechao Meng and Genhao Liang and Xiaofang Zhai and Yalin Lu and Yongqing Li and Lan Chen  and Kehui Wu},
   doi = {10.1088/1674-1056/27/7/076801},
   issn = {1674-1056},
   issue = {7},
   journal = {Chinese Physics B},
   month = {7},
   pages = {076801},
   publisher = {IOP Publishing},
   title = {Growth and transport properties of topological insulator \ce{Bi2Se3} thin film on a ferromagnetic insulating substrate*},
   volume = {27},
   url = {https://iopscience.iop.org/article/10.1088/1674-1056/27/7/076801 https://iopscience.iop.org/article/10.1088/1674-1056/27/7/076801/meta},
   year = {2018}
}

@article{interactivexrdfit,
author = {Lichtensteiger, Céline},
year = {2018},
month = {10},
pages = {1745-1751},
title = {InteractiveXRDFit: a new tool to simulate and fit X-ray diffractograms of oxide thin films and heterostructures},
volume = {51},
journal = {Journal of Applied Crystallography},
doi = {10.1107/S1600576718012840}}

@article{AlegriaLD2014,
  title={Large anomalous Hall effect in ferromagnetic insulator-topological insulator heterostructures},
  author={Alegria, LD and Ji, H and Yao, N and Clarke, JJ and Cava, Robert J and Petta, Jason R},
  journal={Applied Physics Letters},
  volume={105},
  number={5},
  year={2014},
  doi={10.1063/1.4892353},
  publisher={AIP Publishing}}

@article{ChangCZ2013,
  title={Experimental observation of the quantum anomalous Hall effect in a magnetic topological insulator},
  author={Chang, Cui-Zu and Zhang, Jinsong and Feng, Xiao and Shen, Jie and Zhang, Zuocheng and Guo, Minghua and Li, Kang and Ou, Yunbo and Wei, Pang and Wang, Li-Li and others},
  journal={Science},
  volume={340},
  number={6129},
  pages={167--170},
  year={2013},
  publisher={American Association for the Advancement of Science}}

@article{CheckelskyJG2014,
  title={Trajectory of the anomalous Hall effect towards the quantized state in a ferromagnetic topological insulator},
  author={Checkelsky, JG and Yoshimi, R and Tsukazaki, A and Takahashi, KS and Kozuka, Y and Falson, J and Kawasaki, M and Tokura, Y},
  journal={Nature Physics},
  volume={10},
  number={10},
  pages={731--736},
  year={2014},
  doi={},
  publisher={Nature Publishing Group UK London}}

@article{DengY2020,
  title={Quantum anomalous Hall effect in intrinsic magnetic topological insulator MnBi2Te4},
  author={Deng, Yujun and Yu, Yijun and Shi, Meng Zhu and Guo, Zhongxun and Xu, Zihan and Wang, Jing and Chen, Xian Hui and Zhang, Yuanbo},
  journal={Science},
  volume={367},
  number={6480},
  pages={895--900},
  year={2020},
  doi={10.1126/science.aax8156},
  publisher={American Association for the Advancement of Science}}

@article{GongY2019,
  title={Experimental realization of an intrinsic magnetic topological insulator},
  author={Gong, Yan and Guo, Jingwen and Li, Jiaheng and Zhu, Kejing and Liao, Menghan and Liu, Xiaozhi and Zhang, Qinghua and Gu, Lin and Tang, Lin and Feng, Xiao and others},
  journal={Chinese Physics Letters},
  volume={36},
  number={7},
  pages={076801},
  year={2019},
  doi={10.1088/0256-307X/36/7/076801},
  publisher={IOP Publishing}}

@article{HuangH2024,
  title={Magnetic coupling mechanism of the interface in Bi2Se3/La0. 7Sr0. 3MnO3},
  author={Huang, He and Liu, Qiya and Zhang, Min},
  journal={Journal of Alloys and Compounds},
  volume={976},
  pages={173099},
  year={2024},
  doi={10.1016/j.jallcom.2023.173099},
  publisher={Elsevier}}

@article{HwangHY2012,
  title={Emergent phenomena at oxide interfaces},
  author={Hwang, Harold Y and Iwasa, Yoh and Kawasaki, Masashi and Keimer, Bernhard and Nagaosa, Naoto and Tokura, Yoshinori},
  journal={Nature materials},
  volume={11},
  number={2},
  pages={103--113},
  year={2012},
  doi={},
  publisher={Nature Publishing Group UK London}}

@article{KangKT2019,
  title={Synergetic Behavior in 2D Layered Material/Complex Oxide Heterostructures},
  author={Kang, Kyeong Tae and Park, Jeongmin and Suh, Dongseok and Choi, Woo Seok},
  journal={Advanced Materials},
  volume={31},
  number={34},
  pages={1803732},
  year={2019},
  publisher={Wiley Online Library}}

@article{KatmisF2016,
    author = {Katmis, Ferhat and Lauter, Valeria and Nogueira, Flavio and Assaf, Badih and Jamer, Michelle and Wei, Peng and Satpati, Biswarup and Freeland, J. and Eremin, Ilya and Heiman, D. and Jarillo-Herrero, Pablo and Moodera, Jagadeesh},
    year = {2016},
    month = {05},
    pages = {},
    title = {A High-Temperature Ferromagnetic Topological Insulating Phase by Proximity Coupling},
    volume = {533},
    journal = {Nature},
    doi = {10.1038/nature17635}}

@article{KorenG2018,
  title={Magnetic proximity effect of a topological insulator and a ferromagnet in thin-film bilayers of Bi 0.5 Sb 1.5 Te 3 and SrRuO 3},
  author={Koren, Gad},
  journal={Physical Review B},
  volume={97},
  number={5},
  pages={054405},
  year={2018},
  doi={10.1103/PhysRevB.97.054405},
  publisher={APS}}

@article{LiM2017,
  title = {Dirac-electron-mediated magnetic proximity effect in topological insulator/magnetic insulator heterostructures},
  author = {Li, Mingda and Song, Qichen and Zhao, Weiwei and Garlow, Joseph A. and Liu, Te-Huan and Wu, Lijun and Zhu, Yimei and Moodera, Jagadeesh S. and Chan, Moses H. W. and Chen, Gang and Chang, Cui-Zu},
  journal = {Phys. Rev. B},
  volume = {96},
  issue = {20},
  pages = {201301},
  numpages = {5},
  year = {2017},
  month = {Nov},
  publisher = {American Physical Society},
  doi = {10.1103/PhysRevB.96.201301}}

@article{LiQ2024,
  title={Recent progress of MnBi 2 Te 4 epitaxial thin films as a platform for realising the quantum anomalous Hall effect},
  author={Li, Qile and Mo, Sung-Kwan and Edmonds, Mark T},
  journal={Nanoscale},
  volume={16},
  number={30},
  pages={14247--14260},
  year={2024},
  doi={10.1039/D4NR00194J},
  publisher={Royal Society of Chemistry}}

@article{LiuT2020,
  title={Optically Controllable 2D Material/Complex Oxide Heterointerface},
  author={Liu, Tao and Han, Cheng and Xiang, Du and Han, Kun and Ariando, Ariando and Chen, Wei},
  journal={Advanced Science},
  volume={7},
  number={21},
  pages={2002393},
  year={2020},
  publisher={Wiley Online Library}}

@article{LiuY2012,
   author = {Y. Liu and M. Weinert and L. Li},
   doi = {10.1103/PhysRevLett.108.115501},
   issn = {00319007},
   issue = {11},
   journal = {Physical Review Letters},
   month = {3},
   title = {Spiral growth without dislocations: Molecular beam epitaxy of the topological insulator \ce{Bi2Se3} on epitaxial graphene/SiC(0001)},
   volume = {108},
   year = {2012}}

@article{LuQ2022,
  title={Giant tunable spin Hall angle in sputtered Bi2Se3 controlled by an electric field},
  author={Lu, Qi and Li, Ping and Guo, Zhixin and Dong, Guohua and Peng, Bin and Zha, Xi and Min, Tai and Zhou, Ziyao and Liu, Ming},
  journal={Nature communications},
  volume={13},
  number={1},
  pages={1650},
  year={2022},
  doi={10.1038/s41467-022-29281-w},
  publisher={Nature Publishing Group UK London}}

@article{MiaoQ2023,
   author = {Qingqing Miao and Chaoyang Kang and Ye Heng Song and Weifeng Zhang},
   doi = {10.1063/5.0147158},
   issn = {00036951},
   issue = {18},
   journal = {Applied Physics Letters},
   month = {5},
   publisher = {American Institute of Physics Inc.},
   title = {Magnetic proximity effect in the heterostructures of topological insulators and \ce{SrRuO3}},
   volume = {122},
   year = {2023}}

@article{MogiM2015,
   author = {M. Mogi and R. Yoshimi and A. Tsukazaki and K. Yasuda and Y. Kozuka and K. S. Takahashi and M. Kawasaki and Y. Tokura},
   doi = {10.1063/1.4935075/133370},
   issn = {00036951},
   issue = {18},
   journal = {Applied Physics Letters},
   month = {11},
   pages = {56},
   publisher = {American Institute of Physics Inc.},
   title = {Magnetic modulation doping in topological insulators toward higher-temperature quantum anomalous Hall effect},
   volume = {107},
   url = {/aip/apl/article/107/18/182401/133370/Magnetic-modulation-doping-in-topological},
   year = {2015}}

@article{MogiM2017,
  title={A magnetic heterostructure of topological insulators as a candidate for an axion insulator},
  author={Mogi, M and Kawamura, M and Yoshimi, R and Tsukazaki, A and Kozuka, Y and Shirakawa, N and Takahashi, KS and Kawasaki, M and Tokura, Y},
  journal={Nature materials},
  volume={16},
  number={5},
  pages={516--521},
  year={2017},
  doi={10.1038/nmat4855},
  publisher={Nature Publishing Group UK London}}

@article{NittaJ1997,
  title={Gate control of spin-orbit interaction in an inverted I n 0.53 G a 0.47 As/I n 0.52 A l 0.48 as heterostructure},
  author={Nitta, Junsaku and Akazaki, Tatsushi and Takayanagi, Hideaki and Enoki, Takatomo},
  journal={Physical Review Letters},
  volume={78},
  number={7},
  pages={1335},
  year={1997},
  doi={10.1103/PhysRevLett.78.1335},
  publisher={APS}}

@article{OtrokovMM2019,
  title={Prediction and observation of an antiferromagnetic topological insulator},
  author={Otrokov, Mikhail M and Klimovskikh, Ilya I and Bentmann, Hendrik and Estyunin, D and Zeugner, Alexander and Aliev, Ziya S and Ga{\ss}, Sebastian and Wolter, AUB and Koroleva, AV and Shikin, Alexander M and others},
  journal={Nature},
  volume={576},
  number={7787},
  pages={416--422},
  year={2019},
  doi={10.1038/s41586-019-1840-9},
  publisher={Nature Publishing Group UK London}}

@article{PanL2020,
  title={Observation of quantum anomalous Hall effect and exchange interaction in topological insulator/antiferromagnet heterostructure},
  author={Pan, Lei and Grutter, Alexander and Zhang, Peng and Che, Xiaoyu and Nozaki, Tomohiro and Stern, Alex and Street, Mike and Zhang, Bing and Casas, Brian and He, Qing Lin and others},
  journal={Advanced Materials},
  volume={32},
  number={34},
  pages={2001460},
  year={2020},
  doi={10.1002/adma.202001460},
  publisher={Wiley Online Library}}

@article{QiS2020,
  title={Origin of magnetic inhomogeneity in Cr-and V-doped topological insulators},
  author={Qi, Shifei and Liu, Zheng and Chang, Maozhi and Gao, Ruiling and Han, Yulei and Qiao, Zhenhua},
  journal={Physical Review B},
  volume={101},
  number={24},
  pages={241407},
  year={2020},
  doi={10.1103/PhysRevB.101.241407},
  publisher={APS}}

@article{RiddifordL2022,
    author = {Riddiford, Lauren and Grutter, Alexander and Pillsbury, Timothy and Stanley, Max and Hickey, Danielle and Li, Peng and Alem, Nasim and Samarth, Nitin and Suzuki, Yuri},
    year = {2022},
    month = {03},
    pages = {},
    title = {Understanding Signatures of Emergent Magnetism in Topological Insulator/Ferrite Bilayers},
    volume = {128},
    journal = {Physical Review Letters},
    doi = {10.1103/PhysRevLett.128.126802}}

@article{TangC2017,
  title={Above 400-K robust perpendicular ferromagnetic phase in a topological insulator},
  author={Tang, Chi and Chang, Cui-Zu and Zhao, Gejian and Liu, Yawen and Jiang, Zilong and Liu, Chao-Xing and McCartney, Martha R and Smith, David J and Chen, Tingyong and Moodera, Jagadeesh S and others},
  journal={Science advances},
  volume={3},
  number={6},
  pages={e1700307},
  year={2017},
  doi={10.1126/sciadv.1700307},
  publisher={American Association for the Advancement of Science}}

@article{TokuraY2019,
  title={Magnetic topological insulators},
  author={Tokura, Yoshinori and Yasuda, Kenji and Tsukazaki, Atsushi},
  journal={Nature Reviews Physics},
  volume={1},
  number={2},
  pages={126--143},
  year={2019},
  publisher={Nature Publishing Group UK London}}

@article{WangW2019,
  title={Spin chirality fluctuation in two-dimensional ferromagnets with perpendicular magnetic anisotropy},
  author={Wang, Wenbo and Daniels, Matthew W and Liao, Zhaoliang and Zhao, Yifan and Wang, Jun and Koster, Gertjan and Rijnders, Guus and Chang, Cui-Zu and Xiao, Di and Wu, Weida},
  journal={Nature materials},
  volume={18},
  number={10},
  pages={1054--1059},
  year={2019},
  doi={10.1038/s41563-019-0454-9},
  publisher={Nature Publishing Group UK London}}

@article{WatanabeR2019,
  title={Quantum anomalous Hall effect driven by magnetic proximity coupling in all-telluride based heterostructure},
  author={Watanabe, Ryota and Yoshimi, Ryutaro and Kawamura, Minoru and Mogi, Masataka and Tsukazaki, Atsushi and Yu, XZ and Nakajima, Kiyomi and Takahashi, Kei S and Kawasaki, Masashi and Tokura, Yoshinori},
  journal={Applied Physics Letters},
  volume={115},
  number={10},
  year={2019},
  doi={10.1063/1.5111891},
  publisher={AIP Publishing}}

@article{YangXS2020,
  title={Transport properties for Bi2Se3/La0.7Sr0.3MnO3 composites},
  author={Yang, XS and Zhao, T and Liu, QY and Zhao, K and Wei, ZT and Zhao, Y},
  journal={Ceramics International},
  volume={46},
  number={4},
  pages={4748--4753},
  year={2020},
  doi={10.1016/j.ceramint.2019.10.206},
  publisher={Elsevier}}

@article{ZhuS2018,
   author = {Shanna Zhu and Gang Shi and Peng Zhao and Dechao Meng and Genhao Liang and Xiaofang Zhai and Yalin Lu and Yongqing Li and Lan Chen and Kehui Wu},
   doi = {10.1088/1674-1056/27/7/076801},
   issn = {1674-1056},
   issue = {7},
   journal = {Chinese Physics B},
   month = {7},
   pages = {076801},
   publisher = {IOP Publishing},
   title = {Growth and transport properties of topological insulator \ce{Bi2Se3} thin film on a ferromagnetic insulating substrate*},
   volume = {27},
   url = {https://iopscience.iop.org/article/10.1088/1674-1056/27/7/076801 https://iopscience.iop.org/article/10.1088/1674-1056/27/7/076801/meta},
   year = {2018}}

@article{ZhuS2018_2,
   author = {Shanna Zhu and Dechao Meng and Genhao Liang and Gang Shi and Peng Zhao and Peng Cheng and Yongqing Li and Xiaofang Zhai and Yalin Lu and Lan Chen and Kehui Wu},
   doi = {10.1039/C8NR02083C},
   issn = {2040-3372},
   issue = {21},
   journal = {Nanoscale},
   month = {5},
   pages = {10041-10049},
   pmid = {29774918},
   publisher = {The Royal Society of Chemistry},
   title = {Proximity-induced magnetism and an anomalous Hall effect in \ce{Bi2Se3}/\ce{LaCoO3}: a topological insulator/ferromagnetic insulator thin film heterostructure},
   volume = {10},
   url = {https://pubs.rsc.org/en/content/articlehtml/2018/nr/c8nr02083c https://pubs.rsc.org/en/content/articlelanding/2018/nr/c8nr02083c},
   year = {2018}}

@article{KimuraT1996,
    title = {{Interplane Tunneling Magnetoresistance in a Layered Manganite Crystal}},
    year = {1996},
    journal = {Science},
    author = {Kimura, T. and Tomioka, Y. and Kuwahara, H. and Asamitsu, A. and Tamura, M. and Tokura, Y.},
    number = {5293},
    month = {12},
    pages = {1698--1701},
    volume = {274},
    publisher = {American Association for the Advancement of Science},
    url = {/doi/pdf/10.1126/science.274.5293.1698},
    doi = {10.1126/SCIENCE.274.5293.1698},
    issn = {00368075},
    pmid = {8939857}}

@article{BerndtLM2000,
  title={Magnetic anisotropy and strain states of (001) and (110) colossal magnetoresistance thin films},
  author={Berndt, LM and Balbarin, Vincent and Suzuki, Y},
  journal={Applied Physics Letters},
  volume={77},
  number={18},
  pages={2903--2905},
  year={2000},
  doi={10.1063/1.1321733},
  publisher={American Institute of Physics}}

@article{LloydJ2017,
  title={Colossal terahertz magnetoresistance at room temperature in epitaxial La0.7Sr0.3MnO3 nanocomposites and single-phase thin films},
  author={Lloyd-Hughes, James and Mosley, CDW and Jones, SPP and Lees, Martin R and Chen, Aiping and Jia, Quan Xi and Choi, E-M and MacManus-Driscoll, JL},
  journal={Nano Letters},
  volume={17},
  number={4},
  pages={2506--2511},
  year={2017},
  doi={10.1021/acs.nanolett.7b00231},
  publisher={ACS Publications}}

@article{TokuraY2000,
  title={Correlated electrons: science to technology},
  author={Tokura, Yoshinori},
  journal={JSAP Int},
  volume={2},
  pages={12--21},
  year={2000}}

@article{TsengLT2015,
  title={Magnetic properties in $\alpha$-MnO2 doped with alkaline elements},
  author={Tseng, Li-Ting and Lu, Yunhao and Fan, Hai Ming and Wang, Yiren and Luo, Xi and Liu, Tao and Munroe, Paul and Li, Sean and Yi, Jiabao},
  journal={Scientific reports},
  volume={5},
  number={1},
  pages={9094},
  year={2015},
  doi={10.1038/srep09094},
  publisher={Nature Publishing Group UK London}}

@article{IzumiM2001,
  title={Perovskite superlattices as tailored materials of correlated electrons},
  author={Izumi, M and Ogimoto, Y and Konishi, Y and Manako, T and Kawasaki, M and Tokura, Y},
  journal={Materials Science and Engineering: B},
  volume={84},
  number={1-2},
  pages={53--57},
  year={2001},
  doi={10.1016/S0921-5107(01)00569-4},
  publisher={Elsevier}}

@article{ChenYL2009,
  title={Experimental realization of a three-dimensional topological insulator, Bi2Te3},
  author={Chen, YL and Analytis, James G and Chu, J-H and Liu, ZK and Mo, S-K and Qi, Xiao-Liang and Zhang, HJ and Lu, DH and Dai, Xi and Fang, Zhong and others},
  journal={science},
  volume={325},
  number={5937},
  pages={178--181},
  year={2009},
  doi={10.1126/science.1173034},
  publisher={American Association for the Advancement of Science}}

@article{VazquezA2005,
  title={One-step synthesis of Mn3O4 nanoparticles: Structural and magnetic study},
  author={V{\'a}zquez-Olmos, Am{\'e}rica and Red{\'o}n, Roc{\'\i}o and Rodr{\'\i}guez-Gattorno, Geonel and Mata-Zamora, M Esther and Morales-Leal, Francisco and Fern{\'a}ndez-Osorio, Ana L and Saniger, Jos{\'e} M},
  journal={Journal of colloid and interface science},
  volume={291},
  number={1},
  pages={175--180},
  year={2005},
  doi={10.1016/j.jcis.2005.05.005},
  publisher={Elsevier}}

@article{ZhangH2009,
  title={Topological insulators in Bi2Se3, Bi2Te3 and Sb2Te3 with a single Dirac cone on the surface},
  author={Zhang, Haijun and Liu, Chao-Xing and Qi, Xiao-Liang and Dai, Xi and Fang, Zhong and Zhang, Shou-Cheng},
  journal={Nature physics},
  volume={5},
  number={6},
  pages={438--442},
  year={2009},
  doi={10.1038/nphys1270},
  publisher={Nature Publishing Group UK London}}

@article{GonzalezRD2024,
  title={Anisotropic magnetoresistance in altermagnetic MnTe},
  author={Gonzalez Betancourt, Ruben Dario and Zubac, Jan and Geishendorf, Kevin and Ritzinger, Philipp and Ruvica, Barbora and Kotte, Tommy and Zelezny, Jakub and Olejnik, Kamil and Springholz, Gunther and Buchner, Bernd and others},
  journal={npj Spintronics},
  volume={2},
  number={1},
  pages={45},
  year={2024},
  doi={10.1038/s44306-024-00046-z},
  publisher={Nature Publishing Group UK London}}

@article{ChenW2020,
  title={Electronic structure and magnetism of mte2 (m= ti, v, cr, mn, fe, co and ni) monolayers},
  author={Chen, Wei and Zhang, Jian-min and Nie, Yao-zhuang and Xia, Qing-lin and Guo, Guang-hua},
  journal={Journal of Magnetism and Magnetic Materials},
  volume={508},
  pages={166878},
  year={2020},
  doi={10.1016/j.jmmm.2020.166878},
  publisher={Elsevier}}

@article{ValvidaresSM2001,
  title={Inverted hysteresis loops in magnetically coupled bilayers with uniaxial competing anisotropies: Theory and experiments},
  author={Valvidares, SM and Alvarez-Prado, LM and Martin, JI and Alameda, JM},
  journal={Physical Review B},
  volume={64},
  number={13},
  pages={134423},
  year={2001},
  doi={10.1103/PhysRevB.64.134423},
  publisher={APS}}

@article{OSheaMJ1994,
  title={Inverted hysteresis in magnetic systems with interface exchange},
  author={O’Shea, MJ and Al-Sharif, A-L},
  journal={Journal of Applied Physics},
  volume={75},
  number={10},
  pages={6673--6675},
  year={1994},
  doi={10.1063/1.356891},
  publisher={American Institute of Physics}}

@article{GhisingP2020,
  title={Spin inhomogeneities at the interface and inverted hysteresis loop in \ce{La0.7Sr0.3MnO3}/\ce{SrTiO3}},
  author={Ghising, Pramod and Samantaray, B and Hossain, Z},
  journal={Physical Review B},
  volume={101},
  number={2},
  pages={024408},
  year={2020},
  doi={10.1103/PhysRevB.101.024408},
  publisher={APS}}

@article{WeiG2026,
  title={Universal inverted hysteresis induced by interfacial noncollinearity: From metallic exchange-spring bilayers to ferromagnet/topological insulator heterostructures},
  author={Wei, Guodong and Gong, Hanqing and Bai, Ligang and Wang, Hangtian and Zhou, Zhiqin and Xie, Weiran and Huang, Hanyu and Zou, Jin and Zhao, Weisheng and Chen, Yanxue and others},
  journal={Physical Review B},
  volume={113},
  number={10},
  pages={104424},
  year={2026},
  doi={10.1103/v86b-m745},
  publisher={APS}}

@article{BhardwajRG2025,
  title={Canted magnetism and topological spin texture induced in silicon from flexoelectronic proximity effect},
  author={Bhardwaj, Ravindra G and Katailiha, Anand and Lou, Paul C and Beyermann, WP and Kumar, Sandeep},
  journal={Journal of Magnetism and Magnetic Materials},
  volume={618},
  pages={172886},
  year={2025},
  doi={10.1016/j.jmmm.2025.172886},
  publisher={Elsevier}}

@article{HallsteinsenI2016,
  title={Concurrent magnetic and structural reconstructions at the interface of (111)-oriented L a 0.7 S r 0.3 Mn O 3/LaFe O 3},
  author={Hallsteinsen, I and Moreau, M and Grutter, Alexander and Nord, M and Vullum, P-E and Gilbert, Dustin A and Bolstad, T and Grepstad, JK and Holmestad, R and Selbach, SM and others},
  journal={Physical Review B},
  volume={94},
  number={20},
  pages={201115},
  year={2016},
  doi={10.1103/PhysRevB.94.201115},
  publisher={APS}}

@article{KavichJJ2007,
  title={Nanoscale suppression of magnetization at atomically assembled manganite interfaces: XMCD and XRMS measurements},
  author={Kavich, JJ and Warusawithana, MP and Freeland, JW and Ryan, P and Zhai, X and Kodama, RH and Eckstein, JN},
  journal={Physical Review B—Condensed Matter and Materials Physics},
  volume={76},
  number={1},
  pages={014410},
  year={2007},
  doi={10.1103/PhysRevB.76.014410},
  publisher={APS}}

\end{document}

\maketitle

% The thickness of sample S1 was directly measured using cross-sectional TEM. Figure \ref{00} shows a TEM scan displaying all layers of the stack. The topmost layer (light gray color) is identified as \ce{Bi2Te3} with the thickness of about 41~nm. The darker stripes present in the layer, parallel to the substrate surface, indicate the van der Waals gaps between BT's quintuple layers whose thickness is approximately 1~nm. The following interfacial layer does not display an apparent crystallographic ordering. The thickness of the LSMO layer is estimated to be about 5~nm. 
% \begin{figure}
%     \centering
%     \includegraphics[width=0.67\linewidth]{SIfig/SI00.pdf}
%     \caption{Cross-sectional TEM of sample S1 used to approximate the thickness of the layers deposited.}
%     \label{00}
% \end{figure}

\section*{Structural Analysis}
\begin{figure}[H]
    \centering
    \includegraphics[width=0.67\linewidth]{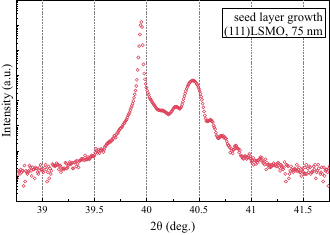}
    \caption{HRXRD of the seed layer heterostructure showing the (111)STO substrate peak and the LSMO peak with the estimated layer thickness of about 75~nm.}
    \label{001}
\end{figure}

Figure \ref{001} shows HRXRD scan of the seed-layer sample. The peak yielding the highest intensity corresponds to the (111)STO substrate, while the (111)LSMO peak is located at $2\theta$~=~40.45°. Based on the distance between the lattice fringes of the LSMO film, the thickness of the film is approximately 75~nm.

\section*{Magnetometry Measurements}
\begin{figure}[H]
    \centering
    \includegraphics[width=0.9\linewidth]{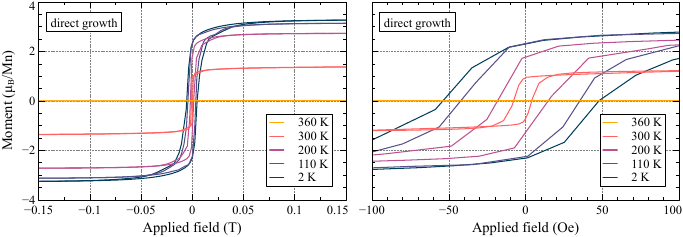}
    \caption{Magnetic hysteresis curves of the direct growth heterostructure collected at a range of temperatures. The right side plot shows a zoomed-in region at applied field $H$~=~$\pm$100~Oe.}
    \label{03}
\end{figure}

\begin{figure}[H]
    \centering
    \includegraphics[width=0.9\linewidth]{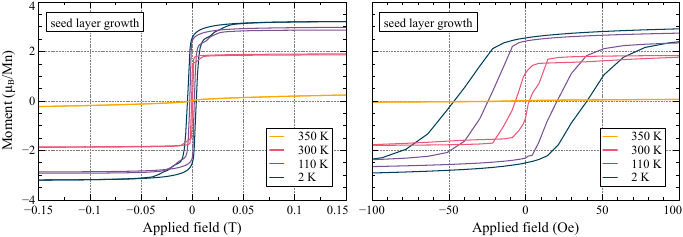}
    \caption{Magnetic hysteresis curves of the seed layer heterostructure collected at a range of temperatures. The right side plot shows a zoomed-in region at applied field $H$~=~$\pm$100~Oe.}
    \label{04}
\end{figure}

\begin{figure}[H]
    \centering
    \includegraphics[width=0.9\linewidth]{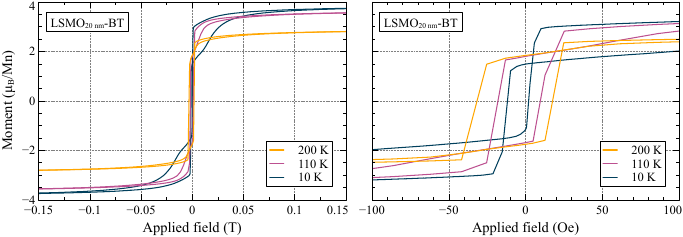}
    \caption{Magnetic hysteresis curves of the heterostructure with thin LSMO layer collected at a range of temperatures. The right side plot shows a zoomed-in region at applied field $H$~=~$\pm$100~Oe.}
    \label{02}
\end{figure}

% \section{X-ray Magnetic Circular Dichroism}

\section*{Polarized Neutron Reflectometry}
The process of building the precise model relies on creating the simplest model expected, which serves as a base for further building up of the fitting parameters. It must be noted that the figure of merit $\chi^2$ does not reflect the accuracy to the real state of the stack, but rather how well the collected data is fit. Therefore, in order to create a model that is reasonably accurate, it is critical to collect preliminary data on the modeled systems. For this study, HRXRD was used to estimate the thickness of the layers through the fitting process in InteractiveXRDFit. Raman spectroscopy was used to confirm the correct stoichiometry of \ce{Bi2Te3}, and bulk magnetometry was used to measure field dependent magnetization in both samples.\newline

In the case of LSMO-BT heterostructures with seeded and direct deposition, the main questions to be answered by PNR are: 1) Does either of deposition approaches lead to observable changes at the interface? 2) Is there a magnetic phase present outside of LSMO?\newline

Model~1 is created to fit the data of the heterostructure with direct deposition. It consists of magnetic LSMO and BT, grown on STO substrate. The starting thickness values are set based on the HRXRD data and equal 79.0~nm and 20.0~nm for LSMO and BT, respectively. Both parameters are set to fit between the range of $\pm$5~nm. LSMO has a magnetic scattering length density component (\textit{rhoM}) initially equal 2.0 and fit between the range of 0.0$\times$10$^{-4}$~nm$^{-2}$ to 5.0$\times$10$^{-4}$~nm$^{-2}$. In addition, the roughness of each layer (\textit{interface}) is set as a fitting parameter between 0.0~nm and 10.0~nm. The initial roughness values are equal 2.0~nm for STO, 5.0~nm for LSMO, and 6.0~nm for BT. The scattering length density values (\textit{rho}) are taken from the NIST Neutron Activation and Scattering Calculator, and their fitting range is set to $\pm$10\%.\newline

\begin{figure}[H]
    \centering
    \includegraphics[width=\linewidth]{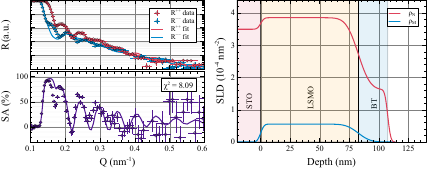}
    \caption{PNR Model 1: basic LSMO and BT stack. Thickness values implemented in the model are based on the HRXRD and XRR data. The LSMO layer is magnetic, while for BT the magnetic fit parameters are fixed to zero. The left panel shows reflectivity data $R$ of both spin channels along with the model fit, as well as the calculated spin asymmetry $SA$. The right panel shows the suggested SLD and mSLD profiles. The colored areas highlight the layers present in the stack.}
    \label{M1}
\end{figure}

The results for Model~1 are summarized in Figure~\ref{M1}. The left panel shows reflectivity data (points) and fit (lines) plotted in the logarithmic scale, as well as the resulting spin asymmetry \textit{SA}. The right panel shows the SLD and mSLD profiles along with the shaded areas that represent the thickness of the layers predicted by the model. The figure of merit $\chi^2$ reaches 8.09, indicating a poor match with the data. It is evident from the \textit{SA} plot that the fit is greatly mismatched at low \textit{Q}. The values achieved for thickness parameters differ noticeably from the initial values and equal 82.6~nm for LSMO and 25.0~nm for BT. Notably, the value obtained for BT resides at the limit of the fitting range. The mSLD of LSMO is equal 0.55$\times$10$^{-4}$~nm$^{-2}$. Such large deviation from the initial value may be a result of the temperature at which the measurements were carried out ($T$~300~K). Finally, the roughness values are 2.3~nm for STO, 8.1~nm for LSMO, and 2.0~nm for BT.\newline

\begin{figure}[H]
    \centering
    \includegraphics[width=\linewidth]{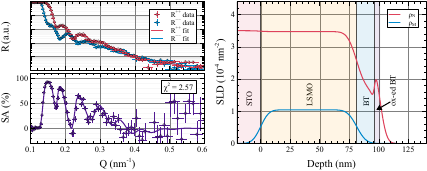}
    \caption{PNR Model 2: LSMO and BT stack with oxidized BT upper layer. Thickness values implemented in the model are based on the HRXRD and XRR data. The LSMO layer is magnetic, while for BT and oxidized BT the magnetic fit parameters are fixed to zero. The left panel shows reflectivity data $R$ of both spin channels along with the model fit, as well as the calculated spin asymmetry $SA$. The right panel shows the suggested SLD and mSLD profiles. The colored areas highlight the layers present in the stack.}
    \label{M2}
\end{figure}

The following Model~2 assumes the presence of the oxidized topmost layer of BT. Oxidation effects are commonly observed in uncapped BT thin films and affect the scattering length density. The density \textit{rho} of the oxidized BT is set to 2.5$\times$10$^{-4}$~nm$^{-2}$ with $\pm$10\% fitting range. The thickness is set to 5.0~nm, and the roughness to 1.0~nm. All parameters of the remaining layers remain unchanged. Figure~\ref{M2} shows the result of Model~2. The fitting is greatly improved and $\chi^2$ reaches 2.57. Notably, the fit appears much more accurate at \textit{Q} below 0.2~nm$^{-1}$. The thickness values are 81.4~nm for LSMO, 15.0~nm for BT, and 4.5~nm for oxidized BT. Once again, the model assumes the thickness of BT at the limit of the fitting range. However, the lower limit is preferred with this model iteration. This is expected, considering that the oxidized BT layer is introduced. Together, they form a layer of 19.5~nm, which is in agreement with the HRXRD data. What is more, mSLD of LSMO is considerably different and reaches 1.0$\times$10$^{-4}$~nm$^{-2}$. Such discrepancy may indicate that there may be additional magnetic component present within the stack.\newline

\begin{figure}[H]
    \centering
    \includegraphics[width=\linewidth]{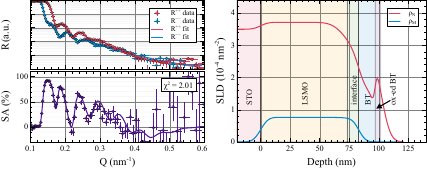}
    \caption{PNR Model 3: including the interfacial phase. The LSMO layer is magnetic, while for the interface, BT, and oxidized BT the magnetic fit parameters are fixed to zero. The left panel shows reflectivity data $R$ of both spin channels along with the model fit, as well as the calculated spin asymmetry $SA$. The right panel shows the suggested SLD and mSLD profiles. The colored areas highlight the layers present in the stack.}
    \label{M3}
\end{figure}

Model~3, shown in Figure~\ref{M3}, adds non-magnetic interfacial layer between LSMO and BT, with the thickness of 5.0~nm fitting between the range from 0.0~nm to 10.0~nm. SLD of the interfacial layer is set to 2.5$\times$10$^{-4}$~nm$^{-2}$ and the fitting range of 0.5-3.0$\times$10$^{-4}$~nm$^{-2}$. With the interfacial layer added, the bottom fitting limit of the BT decreased from 15~nm to 10~nm. The fit is improved and $\chi^2$ reaches 2.01. Thickness values equal 73.2~nm for LSMO, 9.9~nm for the interface, 13.8~nm for BT, and 5.1~nm for oxidized BT. Magnetic scattering length density of LSMO is fit to 0.8$\times$10$^{-4}$~nm$^{-2}$, once more changing considerably between the models.\newline

\begin{figure}[H]
    \centering
    \includegraphics[width=\linewidth]{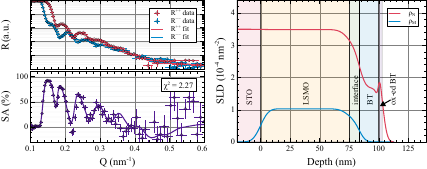}
    \caption{PNR Model 4: magnetic fit parameters of the interfacial phase are set within the model as variables. The LSMO and the interfacial phase layers are magnetic, while for BT and oxidized BT the magnetic fit parameters are fixed to zero. The left panel shows reflectivity data $R$ of both spin channels along with the model fit, as well as the calculated spin asymmetry $SA$. The right panel shows the suggested SLD and mSLD profiles. The colored areas highlight the layers present in the stack.}
    \label{M4}
\end{figure}

Figure~\ref{M4} summarizes the results of Model~4, in which \textit{rhoM} fit parameter is added to the interfacial layer. The initial \textit{rhoM} value is set to 0.0$\times$10$^{-4}$~nm$^{-2}$ and the fitting range is 0.0-1.0$\times$10$^{-4}$~nm$^{-2}$. The model results in a marginally poorer fit with $\chi^2$~=~2.27. The \textit{SA} plot reveals that the fit above $Q$~=~0.2~nm$^{-1}$ misses the features observed in the data such as the negative spin asymmetry at $\sim$0.22~nm$^{-1}$. Notably, \textit{rhoM} of the interfacial layer has the value of 0.59$\times$10$^{-4}$~nm$^{-2}$.\newline

\begin{figure}[H]
    \centering
    \includegraphics[width=\linewidth]{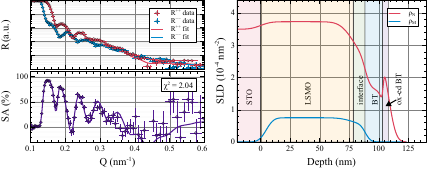}
    \caption{PNR Model 5: based on Model 4, where an additional fit parameter is added for the LSMO layer to account for magnetic dead layer between LSMO and STO. The LSMO and the interfacial phase layers are magnetic, while for BT and oxidized BT the magnetic fit parameters are fixed to zero. The left panel shows reflectivity data $R$ of both spin channels along with the model fit, as well as the calculated spin asymmetry $SA$. The right panel shows the suggested SLD and mSLD profiles. The colored areas highlight the layers present in the stack.}
    \label{M5}
\end{figure}

Model~5 shown in Figure~\ref{M5} adds LSMO magnetic dead layer to fit parameters. The dead layer is set between STO and LSMO. The remaining parameters remain as in the previous model. $\chi^2$ is improved relative to that of Model~4 and equals 2.04. Once again, the interfacial layer yields non-zero \textit{rhoM}~=~0.50$\times$10$^{-4}$~nm$^{-2}$. Importantly, this model fits the negative spin asymmetry at $Q$~=~0.22~nm$^{-1}$ more closely.
The following Model~6 is presented in the main text. It adds the topmost surface layer to account for surface contamination, and yields the best fit with $\chi^2$~=~1.69.\newline

\begin{figure}[H]
    \centering
    \includegraphics[width=\linewidth]{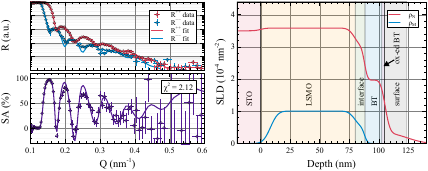}
    \caption{PNR Model 7: Model 6 applied to the data of sample S4. The LSMO and the interfacial phase layers are magnetic, while for the remaining layers the magnetic fit parameters are fixed to zero. The left panel shows reflectivity data $R$ of both spin channels along with the model fit, as well as the calculated spin asymmetry $SA$. The right panel shows the suggested SLD and mSLD profiles. The colored areas highlight the layers present in the stack.}
    \label{M7}
\end{figure}

The following models are applied to the second sample grown with the intermediate tellurium seed layer between LSMO and BT. Firstly, Model~7 is built based on Model~6 of the previous sample. Here, initial thickness parameters are changed to match the results obtained from the HRXRD data for the second sample. The remaining parameters remain the same. $\chi^2$ of Model~7 presented in Figure~\ref{M7} equals 2.12. It is evident from the reflectivity $R$ and spin asymmetry $SA$ plots that the model misses the features at $Q$ above 0.8~nm$^{-1}$, and that the $R^{--}$ data is not well represented within the model. Similar to the first sample, the seed layer sample model results in non-zero \textit{rhoM} at the interface. In addition, SLD of the BT layer is better-defined, indicative of improved layer roughness.\newline

\begin{figure}[H]
    \centering
    \includegraphics[width=\linewidth]{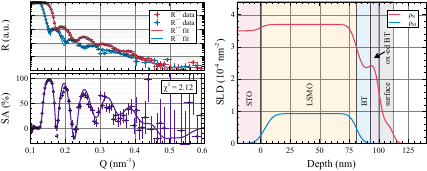}
    \caption{PNR Model 8: based on Model 7, with the interfacial layer removed. The LSMO layer is magnetic, while for the remaining layers the magnetic fit parameters are fixed to zero. The left panel shows reflectivity data $R$ of both spin channels along with the model fit, as well as the calculated spin asymmetry $SA$. The right panel shows the suggested SLD and mSLD profiles. The colored areas highlight the layers present in the stack.}
    \label{M8}
\end{figure}

\begin{figure}
    \centering
    \includegraphics[width=\linewidth]{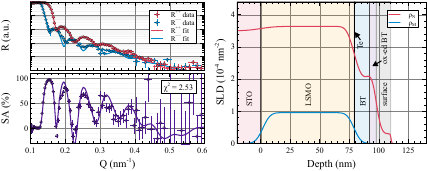}
    \caption{PNR Model 9: based on Model 8, where an intermediate tellurium layer is introduced between LSMO and BT. The LSMO layer is magnetic, while for the remaining layers the magnetic fit parameters are fixed to zero. The left panel shows reflectivity data $R$ of both spin channels along with the model fit, as well as the calculated spin asymmetry $SA$. The right panel shows the suggested SLD and mSLD profiles. The colored areas highlight the layers present in the stack.}
    \label{M9}
\end{figure}

One of the questions to be answered with the PNR analysis is whether the different deposition protocol affects the interface between LSMO and BT. To this end, two following models explore changed conditions between the layers. Model~8 shown in Figure~\ref{M8} removes the intermediate interfacial layer. The resulting fit gives $\chi^2$~=~2.12–the same as that of Model~7. Model~9 shown in Figure~\ref{M9} exchanges the interfacial layer for a separate thin tellurium layer. The initial thickness value is set to 2~nm and ranged between 0-3~nm. The final fit result approximates the thickness of Te to 0.13~nm, and the figure of merit $\chi^2$~=~2.53, indicative of a worse match to the data than two previous models.\newline

\begin{figure}
    \centering
    \includegraphics[width=\linewidth]{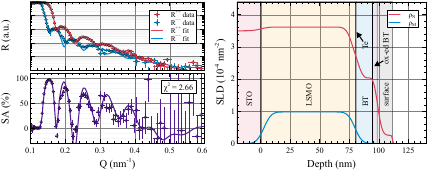}
    \caption{PNR Model 10: based on Model 9, where the tellurium layer has its magnetic fit parameters set as variables. The left panel shows reflectivity data $R$ of both spin channels along with the model fit, as well as the calculated spin asymmetry $SA$. The right panel shows the suggested SLD and mSLD profiles. The colored areas highlight the layers present in the stack.}
    \label{M10}
\end{figure}

A magnetic component \textit{rhoM} is added to the tellurium layer in Model~10 presented in Figure~\ref{M10}. The fitting range is set to 0.00-1.00$\times$10$^{-4}$~nm$^{-2}$, and the final result assumes \textit{rhoM}~=~0.03$\times$10$^{-4}$~nm$^{-2}$. Once again, the figure of merit $\chi^2$ indicates that the model is not improved.
Considering that both models introducing the tellurium layer do not improve the fit, we explore the approach that eliminates intermediate layers between LSMO and BT. The final fit for this dataset is shown in the main text and is based on Model~8, where the LSMO layer is split into two sub-layers.

% \bibliography{references, ref, references_zotero}